\documentclass[10pt,conference]{IEEEtran}

\IEEEoverridecommandlockouts

\usepackage{graphicx}
\usepackage{amsmath,amsxtra, amsfonts, amssymb}
\usepackage{setspace}
\usepackage{times}
\usepackage{epsfig}
\usepackage{subfigure}
\usepackage{latexsym}
\usepackage{cite}
\usepackage{url}
\usepackage{epstopdf}
\usepackage{tikz,pgf,pgfplots}

\newcommand{\nv}{\boldsymbol{n}}

\newcommand{\xv}{\boldsymbol{x}}
\newcommand{\yv}{\boldsymbol{y}}

\newcommand{\Bm}{\boldsymbol{B}}

\newcommand{\Fm}{\boldsymbol{F}}

\newcommand{\Hm}{\boldsymbol{H}}

\newcommand{\Xm}{\boldsymbol{X}}

\newcommand{\Istor}{\mathcal{I}_{sr}}
\newcommand{\Istod}{\mathcal{I}_{sd}}
\newcommand{\Isrtod}{\mathcal{I}_{(s,r) \rightarrow d}}
\newcommand{\Pstor}{\mathcal{P}_{sr}}
\newcommand{\Pstod}{\mathcal{P}_{sd}}
\newcommand{\Psrtod}{\mathcal{P}_{(s,r) \rightarrow d}}
\newcommand{\Pout}{\mathcal{P}}

\newcounter{mytempeqncnt}

\begin{document}
\title{Energy-Efficient Cooperative Protocols for Full-Duplex Relay Channels}

\author{\IEEEauthorblockN{Mohammad~Khafagy\IEEEauthorrefmark{4}, Amr~Ismail\IEEEauthorrefmark{4}, Mohamed-Slim Alouini\IEEEauthorrefmark{4}, and Sonia A\"issa\IEEEauthorrefmark{1} \thanks{This work was funded in part by a grant from King Abdulaziz City for Science and Technology (KACST), Saudi Arabia.}}
\IEEEauthorblockA{\IEEEauthorrefmark{4} King Abdullah University of Science and Technology (KAUST), Thuwal, Makkah Province, Saudi Arabia\\
\IEEEauthorrefmark{1} Institut National de la Recherche Scientifique (INRS-EMT), University of Quebec, Montreal, QC, Canada\\
Email: \{mohammad.khafagy, amrismail.tammam, slim.alouini\}@kaust.edu.sa, aissa@emt.inrs.ca}}
\maketitle

\begin{abstract}
In this work, energy-efficient cooperative protocols are studied for full-duplex relaying (FDR) with loopback interference. In these protocols, relay assistance is only sought under certain conditions on the different link outages to ensure effective cooperation. Recently, an energy-efficient selective decode-and-forward protocol was proposed for FDR, and was shown to outperform existing schemes in terms of outage. Here, we propose an incremental selective decode-and-forward protocol that offers additional power savings, while keeping the same outage performance. We compare the performance of the two protocols in terms of the end-to-end signal-to-noise ratio cumulative distribution function via closed-form expressions. Finally, we corroborate our theoretical results with simulation, and show the relative relay power savings in comparison to non-selective cooperation in which the relay cooperates regardless of channel conditions.
\end{abstract}
\section{Introduction}


Significant research efforts have been directed to multi-hop communications since Van Der Meulen's seminal work on three-terminal communication channels \cite{van1971three} and their capacity theorems due to Cover and El Gamal \cite{cover1979capacity}.
 This is owing to their offered gains in terms of throughput enhancement and coverage extension. In multi-hop communications, an intermediate relay node assists the communication of a source-destination pair by \emph{listening} to source transmissions and \emph{forwarding} to the destination after the necessary processing. For such, the cooperative relay can adopt one of two operation modes depending on how it regulates the simultaneity of its listening and forwarding phases, namely, half-duplex relaying (HDR) or full-duplex relaying (FDR) \cite{COMMAG_TWR_FDR_2009}. In HDR, relay listening and forwarding take place in two time-orthogonal phases, thus introducing a spectral-efficiency loss when compared to direct transmission. On the other hand, FDR supports simultaneous listening/forwarding phases over the same channel resource, and hence it avoids the spectral-efficiency loss in HDR.

Although FDR clearly offers a spectral-efficiency gain by eliminating the known prelog factor from the capacity expressions of relay channels \cite{COMMAG_TWR_FDR_2009}, performance can be significantly degraded in practice from a different perspective. Indeed, since the relay transmits and receives over the same channel resource, an interference link, called loopback or echo interference link, is introduced from the relay transmitter to its receiver \cite{mobicom2011fullduplex,Duarte_Full_Duplex}. Currently, all known isolation and cancellation techniques for FDR, see \cite{mobicom2011fullduplex,Duarte_Full_Duplex,riihonen200904WCNC,riihonen200906TWC,riihonen201109TWC} and the references therein, cannot \emph{perfectly} prevent relay transmissions from leaking towards its receive antenna and causing undesirable interference. In fact, a level of residual self-interference (RSI) persists and adversely affects the effective signal-to-noise ratio (SNR) inside the capacity logarithm. Thus, increasing the relay transmit power does not necessarily help in boosting the end-to-end performance in FDR \cite{riihonen200904WCNC,riihonen200906TWC,riihonen201109TWC}, which defines one challenge in practical FDR channels. Another challenge in FDR lies in exploiting the cooperative diversity that exists in the channel due to the two message replicas received at the destination from the source and relay.  Recent work on FDR with RSI adopted symbol-by-symbol decoding at the destination \cite{riihonen200904WCNC,riihonen200906TWC,riihonen201109TWC}. However, since the relay forwards a delayed version of the source signal, this decoding approach could not exploit the available diversity, and the direct source transmissions were treated as noise to the stronger relay signal.

\begin{figure}[t!]
\centering
\includegraphics[width=0.99\columnwidth]{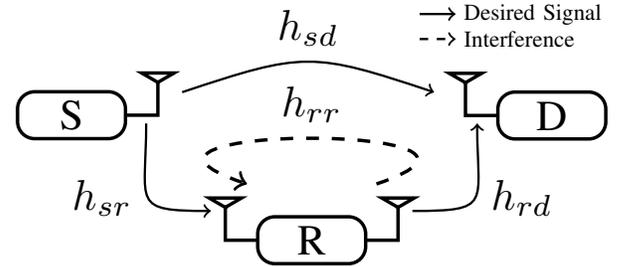}
\caption{System model.}
\label{fig:sys_model}
\end{figure}

In an attempt to tackle the first challenge, we revisit protocols proposed in earlier literature. In \cite{laneman_tse_wornell_2004}, Laneman \textit{et al.} proposed several cooperative protocols for HDR, and studied their outage performance. Among these protocols, selective decode-and-forward (SDF) was proposed in which the relay assists only when the source-relay link is not in outage. Thus, it avoids as much as possible forwarding mere interference to the destination, giving the latter a chance to recover the message from the direct link. Also, incremental decode-and-forward (IDF) was proposed in \cite{laneman_tse_wornell_2004} in which the relay only assists upon the reception of a one-bit feedback from the destination declaring an outage in the direct link. When first proposed, this protocol was primarily sought to alleviate the spectral-efficiency loss introduced in HDR by allowing the destination to possibly rely solely on the direct link as long as it is in a good condition, and thus avoiding the dedication of a time slot for relay forwarding. In recent work \cite{longversion_preprint}, SDF was proposed for further employment in FDR and was shown to outperform existing schemes in terms of outage. With regenerative decode-and-forward relaying adopted, since the relay power directly affects its own channel's outage state, selecting when to cooperate is crucial for the end-to-end performance. In other words, when the relay finds its channel in outage, it is better for it to go into non-cooperative mode. This helps in two directions, namely, saving the relay power, in addition to giving a chance for the destination to attempt decoding from the direct source signal without unnecessary relay interference.

In this work, we tackle the two FDR challenges mentioned earlier. Specifically, for the first challenge, we complement the study in \cite{longversion_preprint} of selective relaying techniques that turn the relay into non-cooperative mode whenever an outage occurs in the source-relay link. Also for the second challenge, block transmission schemes are adopted as in \cite{paulraj2012WCNC,longversion_preprint} in order to exploit the available cooperative diversity. Motivated by the IDF protocol, we propose a protocol that we term incremental selective decode-and-forward (ISDF) for FDR. In addition to the imposed condition on the source-relay link outage, this protocol further demands a one-bit feedback from the destination declaring an outage in the direct link to proceed with cooperation. The ISDF protocol is shown to offer the same outage performance of our previously proposed SDF for FDR, yet yields higher relay power savings. In order to capture the inherent difference in performance between SDF and ISDF, we derive closed-form expressions for the cumulative distribution function (CDF) of the end-to-end SNR. We corroborate our theoretical results with simulations, and show the relative relay power savings in comparison to non-selective cooperation in which the relay cooperates regardless of the channel conditions.

The rest of the paper is organized as follows. In Section \ref{sec:sys_model}, we explain the system model and the notation adopted throughout the paper. We introduce the cooperation policies of the two studied selective relaying protocols in Section \ref{sec:outage} with a brief discussion on their outage performance. In Section \ref{sec:snr}, we derive closed-form expressions for the CDF of the end-to-end SNR for both protocols to capture their relative performance. Numerical simulations are performed and their results are reported in Section \ref{sec:results} to validate our derivations. Finally, we conclude our findings in Section \ref{sec:conc}.
\section{System Model}\label{sec:sys_model}
\subsection{System Model}

We consider the FDR channel depicted in Fig. \ref{fig:sys_model}, where a source (S) communicates with a destination (D) through a direct link with non-negligible gain. A full-duplex relay (R) can possibly assist the S-D communication via a regenerative decode-and-forward (DF) approach. Seeking energy-efficient cooperation, the relay offers assistance only when certain conditions on the different link outages are satisfied as will be shortly highlighted. We denote the channel between node $i \in \{s, r\}$ and node $j \in \{r, d\}$ by $h_{ij}$, and assume that it experiences block fading. Thus, $h_{ij}$ remains constant over one block, and varies independently from one block to another following a circularly symmetric complex Gaussian distribution with zero mean and variance $\pi_{ij}$. Accordingly, the channel envelope is Rayleigh distributed, and thus, the channel gain, $|h_{ij}|^2$, is an exponential random variable with mean $\pi_{ij}$. For notational convenience, we assume that $h_{rr}$ denotes the RSI channel after undergoing all known practical isolation and cancellation techniques (see \cite{mobicom2011fullduplex,Duarte_Full_Duplex,riihonen200904WCNC,riihonen200906TWC,riihonen201109TWC} and the references therein). Also, without loss of generality, all additive white Gaussian noise components are assumed of unit variance.
\subsection{Notation}
Throughout the rest of the paper, we use $x$, $\xv$ and $\Xm$ to respectively denote a scalar quantity, a column vector and a matrix. Also, we use $\Xm^H$ to denote the conjugate transpose of matrix $\Xm$, while $\det\{\Xm\}$ denotes its determinant.
We use $\mathbb{P}\{A\}$ to denote the probability of occurence of an event $A$, while $\mathbb{E}\{.\}$ is used to denote the statistical expectation.
We use $\Gamma_{ij}$ to denote the random variable of the received SNR via the link $h_{ij}$ taking transmit power into account, with $\gamma_{ij}$ being a specific realization of it. Also, $f_{ij}(x)$, $F_{ij}(x)$ and $\overline{F}_{ij}(x)$ are used to respectively denote its probability density function (PDF), CDF and complementary CDF. Moreover, $f_{ij}(x|A)$, $F_{ij}(x|A)$ and $\overline{F}_{ij}(x|A)$ respectively denote its PDF, CDF and complementary CDF conditioned on the occurrence of event $A$. Assuming Gaussian channel inputs and unit bandwidth, we use $\mathcal{I}_{ij}$ to denote the mutual information in the link $h_{ij}$, i.e., $\log_2(1+\Gamma_{ij})$, while $\mathcal{O}_{ij}$ and $\overline{\mathcal{O}}_{ij}$ respectively denote its outage event, i.e., $\mathcal{I}_{ij}<R$ or equivalently $\Gamma_{ij}<\gamma_{th}=2^R-1$, and its complement. Also, we use $\mathcal{P}_{ij}$ to denote the outage probability in the link $h_{ij}$, i.e., $\mathcal{P}_{ij}=\mathbb{P}\{\mathcal{O}_{ij}\}=F_{ij}(\gamma_{th})$, while $\overline{\mathcal{P}}_{ij}=1-\mathcal{P}_{ij}$. 
\subsection{Signal Model}
At time $t$, the source node S transmits $x[t]$ with rate $R=\log_2(1+\gamma_{th})$ bits/s/Hz, and $\mathbb{E}\{|x[t]|^2\}=1$. It is assumed that the outage event dominates the error event. Thus, for a node to be able to attempt decoding the source message, its received SNR should exceed a threshold $\gamma_{th}$. As we mentioned, in the two proposed protocols, certain channel conditions should be satisfied in order to trigger the relay to cooperate. These conditions should at least guarantee that no outage occurs in the S-R link. That is, when the relay enters the cooperative mode, it regenerates the successfully decoded source message taking its processing delay into account. In what follows, we discuss the signal model with and without relay assistance.

\subsubsection{Cooperative Mode}
Concurrently with source transmissions, and taking the relay processing delay $D$ into account, the relay forwards $x[t-D]$ which imposes self-interference on its own received signal. Thus, the received signal at node R is given by:
\begin{eqnarray}
y_r[t]	&=&h_{sr}x[t]+\sqrt{P}h_{rr}x[t-D]+n_r[t],\label{rec_relay}
\end{eqnarray}
where $P$ and $n_r[t]$ denote the relay power and noise, respectively, while the source power is absorbed into both the S-R and S-D channel coefficients without loss of generality.
Hence, the received signal at the destination is given by:
\begin{eqnarray}
y_d[t]	&=&h_{sd}x[t]+\sqrt{P}h_{rd}x[t-D] + n_d[t],\label{rec_dest}
\end{eqnarray}
where $n_d[t]$ denotes the destination noise at time slot $t$.

All channel gains remain constant over a block duration of $L+D$ time slots corresponding to $L$ successive codewords transmitted from the source, in addition to the $D$ time slots delay due to relay processing. Hence, we rewrite (\ref{rec_dest}) in vector form to jointly account for the $L+D$ received signals as:
\begin{eqnarray}
\yv_d=\Hm\xv+\nv_d,\label{rec_sig_ISI}
\end{eqnarray}
where $\yv_d= \left(y_d[1],\ldots,y_d[L+D]\right)^T$, $\xv= \left(x[1],\ldots, x[L]\right)^T$, $\nv_d=\left(n_d[1],\ldots,n_d[L+D]\right)^T$, and
\begin{eqnarray}
\Hm=&h_{sd}\left[ \begin{array}{c} \boldsymbol{I}_L\\ \boldsymbol{0}_{D \times L}\end{array}  \right] + \sqrt{P}h_{rd} \left[ \begin{array}{c} \boldsymbol{0}_{D \times L} \\ \boldsymbol{I}_L\end{array}  \right].
\end{eqnarray}

Assuming complex Gaussian inputs and unit bandwidth, $\mathcal{I}_{ij}$ is given by:
\begin{IEEEeqnarray}[\renewcommand{\IEEEeqnarraymathstyle}{\textstyle}]{rCl}
\mathcal{I}_{ij}~&=&~\log_2\left(1+\Gamma_{ij}\right),
\end{IEEEeqnarray}
where $\Gamma_{sd}=|h_{sd}|^2$ and $\Gamma_{rd}=P|h_{rd}|^2$, while $\Gamma_{sr}=\frac{|h_{sr}|^2}{P|h_{rr}|^2+1}$ denotes the signal-to-interference-plus-noise ratio for the signal arriving at the relay from the source with the RSI effect taken into account.
We use $\Isrtod$ to denote the mutual information per block in the virtual multiple-input single-output (MISO) channel formed by S and R as the transmitter side and by D as the receiver side. Again, assuming complex Gaussian inputs and unit bandwidth, $\Isrtod$ can be readily given by:
\begin{eqnarray}
\Isrtod	=\log_2 \det \left\{\boldsymbol{I}_L+\Hm^H\Hm\right\}=\log_2\prod_{i=1}^L\left(1+\lambda_i\right),\label{information_srd}
\end{eqnarray}
where
\begin{eqnarray}
\Hm^H\Hm	&=& \alpha\boldsymbol{I}_L+\beta{\Bm_L}^D+\beta^*{\Fm_L}^D,
\end{eqnarray}
with $\alpha= |h_{sd}|^2 +P|h_{rd}|^2=\Gamma_{sd}+\Gamma_{rd}$ and $\beta= \sqrt{P} h_{sd}^*h_{rd}$, while $\Bm_L$ ($\Fm_L$) denotes a square backward (forward) shift matrix of size $L$, with ones only on the first subdiagonal (superdiagonal) and zeros elsewhere. Also, $\{\lambda_i\}_{i=1}^L$ denote the eigenvalues of $\Hm^H\Hm$. It can be shown that for $L=mD$, $m\in\mathbb{Z}^+$, the eigenvalues are given as in \cite{longversion_preprint} by:
\begin{IEEEeqnarray}[\renewcommand{\IEEEeqnarraymathstyle}{\textstyle}]{c}
\lambda_{D(i-1)+1:Di}=\alpha+2|\beta|\cos\frac{iD\pi}{L+D},~~~i\in\{1,2,\cdots,m\},\label{eq:general_delay_eigen}
\end{IEEEeqnarray}
where $\lambda_{i:j}$ denotes the set of eigenvalues $\{\lambda_i, \lambda_{i+1},\cdots, \lambda_j\}$.
\subsubsection{Non-Cooperative Mode} In this case, the relay does not assist. Thus, the received signal at the destination at time $t$ is given by:
\begin{eqnarray}
y_d[t]	&=&h_{sd} x[t]+n_d[t].\label{rec_dest_DT}
\end{eqnarray}
For large $\frac{L}{D}$,  in order to keep the same block structure adopted by the no outage case, it is equivalent to use the vector form in (\ref{rec_sig_ISI}) with the value of $P$ set to zero.
\begin{figure*}[t!]
\setcounter{mytempeqncnt}{\value{equation}}
\setcounter{equation}{25}
\begin{eqnarray}
F_{ISDF}(\gamma)		\!\!\!&\approx&\!\!\left\{\begin{array}{ll}
F_{sd}(\gamma)+\frac{P\pi_{rd}\overline{\mathcal{P}}_{sr}\left(F_{rd}(\gamma)-F_{sd}(\gamma)\right)}{(P\pi_{rd}-\pi_{sd})},				 \!\!\!&0<\gamma<\gamma_{th},\\
F_{sd}(\gamma)+\frac{P\pi_{rd}\overline{\mathcal{P}}_{sr}\left(\frac{\mathcal{P}_{sd}}{\mathcal{P}_{rd}}-1\right)\overline{F}_{rd}(\gamma)}{(P\pi_{rd}-\pi_{sd})}	 ,\!\!\!&\gamma_{th}<\gamma<\infty,\end{array}\right.\label{CDF_snr_ISDF}\\
f_{ISDF}(\gamma)		\!\!\!&\approx&\!\!\left\{\begin{array}{ll}
f_{sd}(\gamma)+\frac{P\pi_{rd}\overline{\mathcal{P}}_{sr}\left(f_{rd}(\gamma)-f_{sd}(\gamma)\right)}{(P\pi_{rd}-\pi_{sd})},				 &0<\gamma<\gamma_{th},\\
f_{sd}(\gamma)-\frac{P\pi_{rd}\overline{\mathcal{P}}_{sr}\left(\frac{\mathcal{P}_{sd}}{\mathcal{P}_{rd}}-1\right)f_{rd}(\gamma)}{(P\pi_{rd}-\pi_{sd})}	 ,&\gamma_{th}<\gamma<\infty.\end{array}\right.\label{PDF_snr_ISDF}
\end{eqnarray}
\setcounter{equation}{\value{mytempeqncnt}}
\hrulefill
\end{figure*}
\section{Cooperation Policies and Outage Performance}\label{sec:outage}
\subsection{Selective DF (SDF) Cooperation}
In this part, we briefly explain the cooperation policy and the outage performance of SDF as proposed in \cite{longversion_preprint}. In SDF, cooperation takes place only when the S-R link is not in outage. Thus, an outage is declared when one of two events occurs:
1) the S-R link is in outage, hence no cooperation takes place, while the S-D link goes into an outage state, or 2) the S-R link is not in outage, thus relay assistance is available, but an outage occurs in the MISO channel. This can be more formally defined, as given in \cite{longversion_preprint}, as:
\begin{eqnarray}
\Pout_{SDF}=\Pstor\Pstod+\overline{\mathcal{P}}_{sr}\Psrtod,\label{P_out}
\end{eqnarray}
where
\begin{IEEEeqnarray}[\renewcommand{\IEEEeqnarraymathstyle}{\textstyle}]{rc;l}
\Pstor 	=~&\mathbb{P}\left\{\Istor<R\right\} &=1-\frac{\pi_{sr}e^{-\frac{\gamma_{th}}{\pi_{sr}}}}{\gamma_{th}P\pi_{rr}+\pi_{sr}}\label{P_out_sr},\\
\Pstod	=~&\mathbb{P}\left\{\Istod<R\right\} &=1-e^{-\frac{\gamma_{th}}{\pi_{sd}}}\label{P_out_sd}.
\end{IEEEeqnarray}
while
\begin{IEEEeqnarray}[\renewcommand{\IEEEeqnarraymathstyle}{\textstyle}]{rCl}
\Psrtod	&=&\mathbb{P}\left\{\frac{1}{L+D}\;\Isrtod<\frac{L}{L+D}R\right\}\nonumber\\
		&\approx& \mathbb{P}\left\{\alpha<\gamma_{th}\right\}.\label{P_out_MISO}
\end{IEEEeqnarray}
In \cite{longversion_preprint}, and as given by (\ref{P_out_MISO}), it was shown that the outage probability in the formed MISO channel can be approximated as the probability of $\alpha$ falling below $\gamma_{th}$. This interesting result means that the MISO channel can be equivalently viewed as a point-to-point single-antenna channel with an effective SNR $\Gamma_{eff}\approx\alpha=\Gamma_{sd}+\Gamma_{rd}$. Since $\Gamma_{sd}$ and $\Gamma_{rd}$ are independent exponential random variables with mean parameters $\pi_{sd}$ and $P\pi_{rd}$, respectively, $\alpha$ is a hypoexponential random variable with two rate parameters, $\frac{1}{\pi_{sd}}$ and $\frac{1}{P\pi_{rd}}$. Thus, the CDF of the SNR in the MISO channel, denoted as $F_{(s,r)\rightarrow d}(\gamma)$, is approximately equal to the hypoexponential CDF given in \cite[Eq. (5.9)]{ross-probmod} as:
\begin{IEEEeqnarray}[\renewcommand{\IEEEeqnarraymathstyle}{\textstyle}]{rcl}
F_{(s,r)\rightarrow d}(\gamma) \;&\approx\mathbb{P}\left\{\alpha<\gamma\right\}~	&=\! 	 1-\frac{P\pi_{rd}e^{-\frac{\gamma}{P\pi_{rd}}}-\pi_{sd}e^{-\frac{\gamma}{\pi_{sd}}}}{P\pi_{rd}-\pi_{sd}}.\;\;\;\;\label{cdf_alpha}
\end{IEEEeqnarray}
Therefore, according to the SDF cooperation policy, the received effective SNR at the destination, denoted by $\Gamma$, can be approximately given by:
\begin{eqnarray}
\Gamma &\approx&\left\{\begin{array}{lc}
\Gamma_{sd},&\mathcal{O}_{sr},\\
\Gamma_{sd}+\gamma_{rd},&\overline{\mathcal{O}}_{sr}.
\end{array}
\right.\label{SNR_profile_SDF}
\end{eqnarray}
\subsection{Incremental Selective DF (ISDF) Cooperation}
Here, we propose a different protocol that seeks further relay power savings. We call this protocol as incremental selective DF (ISDF). The selective part comes from the same previously mentioned relay selectivity in forwarding the source message depending on the outage state of the S-R link. However, when outage performance is our main concern, the relay does not need to always forward when it successfully decodes. Instead, relay assistance becomes necessary only upon the reception of a one-bit feedback from the destination at the beginning of the block declaring an outage and asking for assistance. 

Therefore, an outage occurs in the channel under consideration when one of two events occurs: 1) the S-D link goes in outage while the S-R link is in outage. In this case, the relay is unable to assist and outage occurs with probability $\mathcal{P}_{sr}\mathcal{P}_{sd}$ due to the independence of channel fading coefficients, or 2) the S-D link goes in outage while the S-R link is not in outage, but the cooperative MISO channel is in outage. Hence, this  outage event represents an intersection of three events. We know that the event of no outage in the S-R link is independent of the S-D and the MISO channel outage events. Also, we know that cooperation cannot decrease the mutual information, and thus, the outage capacity of the MISO channel is at least equal to that of the S-D link. Hence, the intersection of the outage events in the S-D and the MISO channel is the outage event in the MISO channel itself. Therefore, the probability of this event is equal to $\overline{\mathcal{P}}_{sr}\mathcal{P}_{(s,r)\rightarrow d}$.
Thus, the outage probability in the channel under consideration is given by
\begin{eqnarray}
\mathcal{P}_{ISDF} &=& \mathcal{P}_{sr}\mathcal{P}_{sd} + \overline{\mathcal{P}}_{sr}\mathcal{P}_{(s,r)\rightarrow d}\label{P_out_ISDF},
\end{eqnarray}
while the received SNR profile is approximately given by
\begin{eqnarray}
\Gamma &\approx&\left\{\begin{array}{lc}
\Gamma_{sd},&\overline{\mathcal{O}}_{sd}\cup(\mathcal{O}_{sr}\cap\mathcal{O}_{sd}),\\
\Gamma_{sd}+\Gamma_{rd},&\overline{\mathcal{O}}_{sr}\cap\mathcal{O}_{sd}.
\end{array}
\right.\label{SNR_profile_ISDF}
\end{eqnarray}
Clearly, this outage probability in (\ref{P_out_ISDF}) is exactly equal to that of the SDF protocol given in (\ref{P_out}). Although SDF and ISDF have different cooperation policies, and accordingly different SNR profiles as given in (\ref{SNR_profile_SDF}) and (\ref{SNR_profile_ISDF}), they do have exactly the same outage events, and hence the same outage performance. Since we can easily expect that SDF offers higher performance due to the more cooperation it offers, outage analysis does not suffice to capture and distinguish their relative performance.
\section{SNR Performance}\label{sec:snr}
Due to the selective nature of the adopted cooperation policies, the received SNR at the destination and its probability distribution varies depending on the channel outage states.
Now, we analyze the SNR performance of each of the two protocols.
\subsection{ISDF}
Let $\{A_i\}_{i=1}^4$ denote the intersection of outage events which are defined, along with their probabilities, as follows:
\begin{eqnarray}
\begin{array}{rcl}
A_1 &\triangleq\mathcal{O}_{sr}\cap\mathcal{O}_{sd},				& \mathbb{P}\{A_1\}=\mathcal{P}_{sr}\mathcal{P}_{sd},\\
A_2 &\triangleq\overline{\mathcal{O}}_{sr}\cap\mathcal{O}_{sd},		& \mathbb{P}\{A_2\}=\overline{\mathcal{P}}_{sr}\mathcal{P}_{sd},\\
A_3 &\triangleq\mathcal{O}_{sr}\cap\overline{\mathcal{O}}_{sd},		& \mathbb{P}\{A_3\}=\mathcal{P}_{sr}\overline{\mathcal{P}}_{sd},\\
A_4 &\triangleq\overline{\mathcal{O}}_{sr}\cap\overline{\mathcal{O}}_{sd},	& \mathbb{P}\{A_4\}=\overline{\mathcal{P}}_{sr}\overline{\mathcal{P}}_{sd}.
\end{array}\label{partition_ISDF}
\end{eqnarray}
Clearly, $\{A_i\}_{i=1}^4$ are mutually exclusive events that jointly span the whole sample space, and hence, they form a partitioned space. Therefore, we can use the total probability theorem to get the distribution of $\Gamma$ as follows:
\begin{eqnarray}
F_{ISDF}(\gamma)=\sum_{i=1}^4F(\gamma|A_i)\mathbb{P}\{A_i\},\label{tot_prob_theorem_ISDF}
\end{eqnarray}
where
\begin{eqnarray}
f(\gamma|A_1)					&=&\left\{\begin{array}{lr}\frac{f_{sd}(\gamma)}{\mathcal{P}_{sd}},		 &0<\gamma<\gamma_{th},		\\0,		&\;\;\;\;\;\;\text{elsewhere,}	\end{array}\right.\nonumber\\
f(\gamma|A_3)=f(\gamma|A_4)		&=&\left\{\begin{array}{lr}\frac{f_{sd}(\gamma)}{1-\mathcal{P}_{sd}},	 &\gamma_{th}<\gamma<\infty,		\\0,		&\text{elsewhere,}\end{array}\right.\nonumber
\end{eqnarray}
and thus,
\begin{eqnarray}
F(\gamma|A_1)					 \!\!\!&\!\!=\!\!&\!\!\left\{\begin{array}{lr}\frac{F_{sd}(\gamma)}{\mathcal{P}_{sd}},						 \!\!\!&0<\gamma<\gamma_{th},\\1,&\;\;\;\;\;\;\;\;\;\;\;\;\text{elsewhere,}\end{array}\right.\label{F_given_A1}\\
F(\gamma|A_3)=F(\gamma|A_4)	 \!\!\!&\!\!=\!\!&\!\!\left\{\begin{array}{lr}\frac{F_{sd}(\gamma)-\mathcal{P}_{sd}}{1-\mathcal{P}_{sd}},		 \!\!\!&\gamma_{th}<\gamma<\infty,\\0,&\text{elsewhere.}\end{array}\right.\label{F_given_A3}
\end{eqnarray}
Given $A_2$, $\Gamma_{sd}$ is confined to the range $0<\gamma_{sd}<\gamma_{th}$, thus having the distribution:
\begin{eqnarray}
f(\gamma_{sd}|A_2)					&=&\left\{\begin{array}{lr}\frac{f_{sd}(\gamma_{sd})}{\mathcal{P}_{sd}},		 &0<\gamma_{sd}<\gamma_{th},\\0,&\;\;\;\;\;\;\;\;\;\;\;\;\;\text{elsewhere,}\end{array}\right.\nonumber
\end{eqnarray}
while $\Gamma_{rd}$ remains as an exponential random variable with mean parameter $P\pi_{rd}$ due to independence. Thus, the CDF of $\gamma$ conditioned on $A_2$ is given by:
\begin{eqnarray}
F(\gamma|A_2)					&=&\mathbb{P}\{\Gamma_{sd}+\Gamma_{rd}<\gamma|A_2\}\nonumber\\
							&=&\left\{\begin{array}{ll}F_1(\gamma),		 &0<\gamma<\gamma_{th},\\F_2(\gamma),&\gamma_{th}<\gamma<\infty,\end{array}\right.\label{F_given_A2}
\end{eqnarray}
where
\begin{eqnarray}
F_1(\gamma)	&=&\int_{\gamma_{sd}=0}^{\gamma}f(\gamma_{sd}|A_2)I(\gamma,\gamma_{sd})d\gamma_{sd},\nonumber\\
F_2(\gamma)	 &=&\int_{\gamma_{sd}=0}^{\gamma_{th}}f(\gamma_{sd}|A_2)I(\gamma,\gamma_{sd})d\gamma_{sd},\nonumber
\end{eqnarray}
and
\begin{eqnarray}
I(\gamma,\gamma_{sd})	 =\int_{\gamma_{rd}=0}^{\gamma-\gamma_{sd}}f(\gamma_{rd}|A_2)d\gamma_{rd}=1-e^{-\frac{\gamma-\gamma_{sd}}{P\pi_{rd}}}.\nonumber
\end{eqnarray}
Thus,
\begin{eqnarray}
\!\!\!\!F_1(\gamma)	 \!\!&\!=\!&\frac{F_{sd}(\gamma)}{\mathcal{P}_{sd}}+\frac{P\pi_{rd}\left(F_{rd}(\gamma)-F_{sd}(\gamma)\right)}{\mathcal{P}_{sd}(P\pi_{rd}-\pi_{sd})},\label{F_1}\\
\!\!\!\!F_2(\gamma)	 \!\!&\!=\!&1+\frac{P\pi_{rd}\left(1-e^{-\frac{(P\pi_{rd}-\pi_{sd})\gamma_{th}}{P\pi_{rd}\pi_{sd}}}\right)\overline{F}_{rd}(\gamma)}{\mathcal{P}_{sd}(P\pi_{rd}-\pi_{sd})},\label{F_2}
\end{eqnarray}
where
\begin{eqnarray}
F_{rd}(\gamma)=1-e^{-\frac{\gamma}{P\pi_{rd}}}.\label{F_rd}
\end{eqnarray}

Substituting (\ref{partition_ISDF})-(\ref{F_rd}) into (\ref{tot_prob_theorem_ISDF}), the CDF and the PDF of the end-to-end SNR can be obtained as given by (\ref{CDF_snr_ISDF}) and (\ref{PDF_snr_ISDF}) at the top of this page, respectively. \setcounter{equation}{27}
Also, the average SNR in ISDF is given by
\begin{eqnarray}
\overline{\gamma}_{ISDF}	
						\approx	\int_0^\infty\gamma f_{ISDF}(\gamma)d\gamma=\pi_{sd}+P\pi_{rd}\overline{\mathcal{P}}_{sr}\mathcal{P}_{sd}.
\end{eqnarray}
\subsection{SDF}
\begin{figure*}[t!]
\setcounter{mytempeqncnt}{\value{equation}}
\setcounter{equation}{29}
\begin{eqnarray}
F_{SDF}(\gamma)		\!\!\!&\approx&\!\! 1-\frac{(P\pi_{rd}\mathcal{P}_{sr}-\pi_{sd})\overline{F}_{sd}-(P\pi_{rd}\mathcal{P}_{sr}-P\pi_{rd})\overline{F}_{rd}}{P\pi_{rd}-\pi_{sd}},\label{CDF_snr_SDF}\\
f_{SDF}(\gamma)		\!\!\!&\approx&\!\! \frac{(P\pi_{rd}\mathcal{P}_{sr}-\pi_{sd})f_{sd}(\gamma)-(P\pi_{rd}\mathcal{P}_{sr}-P\pi_{rd})f_{rd}(\gamma)}{(P\pi_{rd}-\pi_{sd})},\label{PDF_snr_SDF}
\end{eqnarray}
\setcounter{equation}{\value{mytempeqncnt}}
\hrulefill
\end{figure*}
In SDF, the relay cooperates regardless of the outage state of the S-D link. Thus, the CDF of the end-to-end SNR can be expressed as:
\begin{eqnarray}
F_{SDF}(\gamma)&=&F(\gamma|\mathcal{O}_{sr})\mathcal{P}_{sr}+F(\gamma|\overline{\mathcal{O}}_{sr})\overline{\mathcal{P}}_{sr}\nonumber\\
			   &\approx& F_{sd}(\gamma)\mathcal{P}_{sr}+F_{(s,r)\rightarrow d}(\gamma)\overline{\mathcal{P}}_{sr}.\label{tot_prob_theorem_SDF}
\end{eqnarray}

Substituting (\ref{cdf_alpha}) in (\ref{tot_prob_theorem_SDF}) and after some manipulations, the CDF and PDF of the end-to-end SNR can be given by (\ref{CDF_snr_SDF}) and (\ref{PDF_snr_SDF}) at the top of the next page. Also, the average SNR in SDF is given by
\setcounter{equation}{31}
\begin{eqnarray}
\overline{\gamma}_{SDF}	\approx		&\displaystyle\int_0^\infty\gamma f_{SDF}(\gamma)d\gamma=\pi_{sd}+P\pi_{rd}\overline{\mathcal{P}}_{sr}.			
\end{eqnarray}
\section{Numerical Results}\label{sec:results}
In this section, we focus on the numerical evaluation of the relative SNR performance in the proposed ISDF scheme compared to that of the SDF scheme proposed in \cite{longversion_preprint}. Since both protocols have identical outage performance, the interested reader is referred to \cite{longversion_preprint} for detailed outage analysis.
\subsection{Simulation Setup}
We generate $10^6$ sets of channel realizations according to the channel model described in previous sections for $L=20$ symbols per block and $D=2$ symbols, i.e., $m=10$. In each set of channel realizations, we exactly calculate $\mathcal{I}_{(s,r)\rightarrow d}$ as given in (\ref{information_srd}) without undergoing any approximations. The relay power, $P$, is set to unity in each set of channel realizations that trigger relay cooperation, while we set its value to zero otherwise. We define the effective end-to-end SNR per symbol in the equivalent single-antenna channel as
\begin{eqnarray}
\Gamma_{eff}=2^{\frac{\mathcal{I}_{(s,r)\rightarrow d}}{L}}-1.\label{Gamma_eff}
\end{eqnarray}
This comes from the equivalence we mentioned earlier between the actual virtual MISO channel and another point-to-point single-antenna channel that can be carefully explained as follows. In the virtual MISO channel, noting that $\mathcal{I}_{(s,r)\rightarrow d}$ denotes the exact mutual information per block, the mutual information per symbol time can be clearly obtained as $\frac{1}{L+D}\mathcal{I}_{(s,r)\rightarrow d}$ since the mutual information per block of $L$ symbols actually spans $L+D$ symbol times. On the other hand, a prelog factor of $\frac{L}{L+D}$ is accordingly introduced in the equivalent single-antenna channel, and hence the mutual information per symbol time is given in terms of $\Gamma_{eff}$ as $\frac{L}{L+D}\log_2\left(1+\Gamma_{eff}\right)$. Thus,
\begin{eqnarray}
\frac{1}{L+D}\mathcal{I}_{(s,r)\rightarrow d}&=&\frac{L}{L+D}\log_2\left(1+\Gamma_{eff}\right),\nonumber
\end{eqnarray}
which yields the expression in (\ref{Gamma_eff}). We compare the average of this $\Gamma_{eff}$ and its empirical probability distribution with those theoretical results obtained in previous sections. For all the figures we present, we include all used simulation parameters in their caption.
\subsection{SNR Performance}
\subsubsection{CDF}
\begin{figure}[t!]
\centering
\includegraphics[width=0.99\columnwidth]{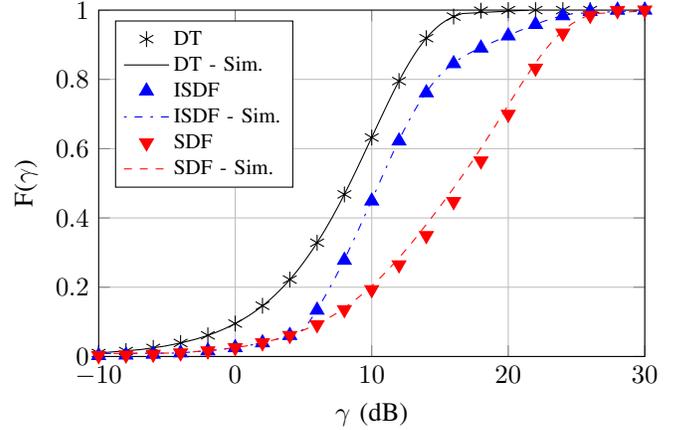}
\caption{CDF of the end-to-end SNR, $F(\gamma)$, for $\pi_{sd}=10$ dB, $\pi_{sr}=20$ dB, $\pi_{rr}=10$ dB, $\pi_{rd}=20$ dB,  $\gamma_{th}=5$ dB, and $R=\log_2(1+\gamma_{th})\approx 2$ bit/s/Hz.}
\label{fig:figure6}
\end{figure}

In Fig. \ref{fig:figure6}, we compare the empirical CDF with that obtained from the derived expressions of the end-to-end SNR via three schemes, namely, (i) non-cooperative direct transmission (DT), (ii) ISDF and (iii) SDF. As depicted in Fig. \ref{fig:figure6}, the CDF of ISDF lies between DT and SDF, and the degree of proximity from either DT or SDF performance depends on the average direct link gain. Specifically, when the direct link has a strong gain, the ISDF yields similar performance to that of DT. This happens for the reason that ISDF does not activate the relay for cooperation as long the destination can retrieve the source message while solely relying on the direct link. On the other hand, when the direct link gain is weak, the performance of ISDF overlaps with that of SDF. This is due to the fact that the destination cannot decode from the source directly, leading to always triggering the relay to cooperate whenever it can decode which is basically what SDF does. We can also notice that the curves of SDF and ISDF overlap till reaching $\gamma_{th}=5$ dB. This confirms what was reached earlier in the outage analysis section, i.e., that both protocols yield the same outage performance.
\subsubsection{Average SNR}
\begin{figure}[t!]
\centering
\includegraphics[width=0.99\columnwidth]{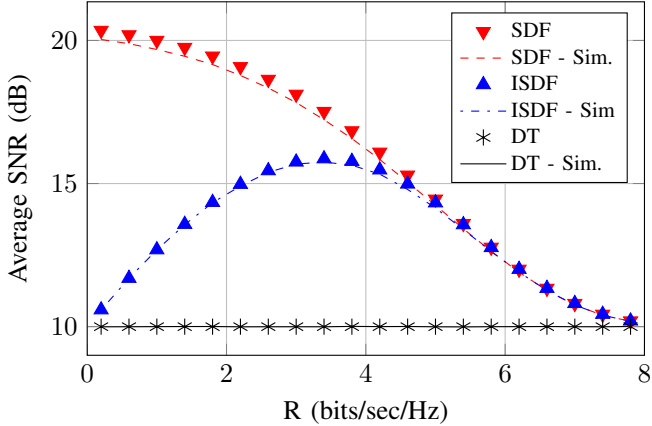}
\caption{Average end-to-end SNR, for $\pi_{sd}=10$ dB, $\pi_{sr}=20$ dB, $\pi_{rr}=10$ dB, and $\pi_{rd}=20$ dB.}
\label{fig:figure7}
\end{figure}

In Fig. \ref{fig:figure7}, we plot the average end-to-end SNR. As we can expect, the average end-to-end SNR of SDF steadily decreases with the increase in source rate. The reason is that, as the source increases its information rate, the relay ability of properly decoding the source message decreases since the S-R link outage occurs more frequently. Thus, no cooperation takes place at that time and the destination receives only from the source. On the other hand, ISDF gives a different performance trend, in which the end-to-end SNR increases starting from the value of the average direct link gain. After some point, it starts its decreasing trend and meets that of SDF till both reach the average direct link gain value at high rates. This is due to the fact that, at very low rates, the destination can anyway decode when solely receiving via the direct link. As the rate increases, the direct link starts to fail more frequently, thus activating the relay cooperation when no outage occurs in the S-R link. As we further increase the rate, outage in the S-R link takes place with higher probability and the relay becomes unable to assist regardless of the direct link's outage state.
\begin{figure}[t!]
\centering
\includegraphics[width=0.99\columnwidth]{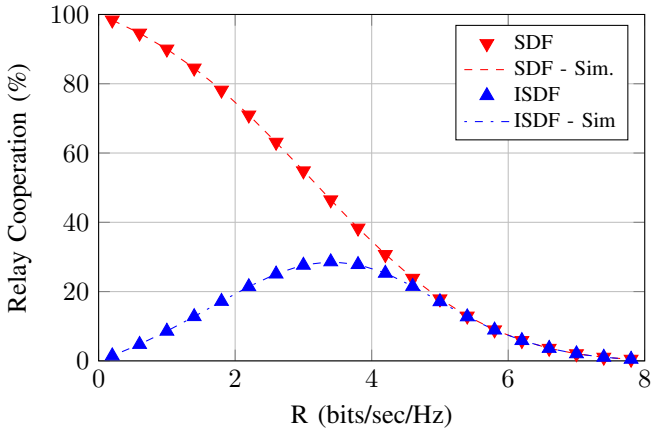}
\caption{Percentage of relay cooperation/power expenditure, for $\pi_{sd}=10$ dB, $\pi_{sr}=20$ dB, $\pi_{rr}=10$ dB, and $\pi_{rd}=20$ dB.}
\label{fig:figure8}
\end{figure}
\subsubsection{Relative Relay Power Expenditure}
In Fig. \ref{fig:figure8}, we plot the percentage of relay power expenditure/relay cooperation obtained via simulation for both schemes. We also compare them with the theoretic values of $(1-\mathcal{P}_{sr})\times100~\%$ and $(1-\mathcal{P}_{sr})\mathcal{P}_{sd}\times100~\%$ for SDF and ISDF, respectively. Over the channel configuration indicated in Fig. \ref{fig:figure8}, ISDF is shown to be more energy-efficient to employ than SDF below $R\approx5$ bits/sec/Hz as $\mathcal{P}_{sd}<1$, since it yields the same outage performance, yet offers significant power savings. This clearly comes at the expense of a lower SNR profile in the no-outage events for ISDF when compared to SDF as noticed earlier from Fig. \ref{fig:figure7}. For rates beyond this value, the direct link falls into outage almost surely, i.e., $\mathcal{P}_{sd}\approx1$. Hence, the relay assists in both protocols whenever it can successfully decode, yielding similar performance for SDF and ISDF as shown in Fig. \ref{fig:figure7} and Fig. \ref{fig:figure8}.

\section{Conclusion}\label{sec:conc}
Energy-efficient cooperative protocols were studied for the full-duplex relay channel with loopback interference. We proposed an incremental selective decode-and-forward protocol that offers additional power savings to recently proposed selective decode-and-forward full-duplex cooperation while maintaining the same outage performance. To capture their inherent performance difference, we derived closed-form expressions for the end-to-end signal-to-noise ratio cumulative distribution function for the two protocols, and showed their well-matching behavior with simulation. Unlike existing FDR protocols where relay cooperation is consistently offered, the studied protocols \emph{selectively} activate the relay for full-duplex cooperation based on the outage states of the different links. Beside the energy efficiency offered by selective cooperation, it alleviates the known adverse effect of the loopback interference in FDR by switching the relay to non-cooperative mode when its level prohibits proper decoding. Also, with block transmission adopted, the proposed protocols could exploit the signal diversity available via the direct link, making them of practical interest for scenarios even beyond coverage extension. 

\bibliographystyle{IEEEbib}
\bibliography{IEEEabrv,ref}

\begin{thebibliography}{10}

\bibitem{van1971three}
E.~C. Van Der~Meulen,
\newblock ``Three-terminal communication channels,''
\newblock {\em Advances in Applied Probability}, pp. 120--154, 1971.

\bibitem{cover1979capacity}
T.~Cover and A.~El Gamal,
\newblock ``Capacity theorems for the relay channel,''
\newblock {\em {IEEE} Trans. Inf. Theory}, vol. 25, no. 5, pp. 572--584, Sept.
  1979.

\bibitem{COMMAG_TWR_FDR_2009}
H.~Ju, E.~Oh, and D.~Hong,
\newblock ``Catching resource-devouring worms in next-generation wireless relay
  systems: Two-way relay and full-duplex relay,''
\newblock {\em {IEEE} Commun. Mag.}, vol. 47, no. 9, pp. 58--65, Sept. 2009.

\bibitem{mobicom2011fullduplex}
M.~Jain, J.~Choi, T.~Kim, D.~Bharadia, S.~Seth, K.~Srinivasan, P.~Levis,
  S.~Katti, and P.~Sinha,
\newblock ``Practical, real-time, full duplex wireless,''
\newblock in {\em Proc. ACM MobiCom'11}, Las Vegas, NV, Sept. 2011.

\bibitem{Duarte_Full_Duplex}
M.~Duarte and A.~Sabharwal,
\newblock ``Full-duplex wireless communications using off-the-shelf radios:
  Feasibility and first results,''
\newblock in {\em Proc. ASILOMAR'10}, Pacific Grove, CA, Nov. 2010.

\bibitem{riihonen200904WCNC}
T.~Riihonen, S.~Werner, R.~Wichman, and J.~Hamalainen,
\newblock ``Outage probabilities in infrastructure-based single-frequency relay
  links,''
\newblock in {\em Proc. IEEE WCNC'09}, Budapest, Hungary, Apr. 2009.

\bibitem{riihonen200906TWC}
T.~Riihonen, S.~Werner, and R.~Wichman,
\newblock ``Optimized gain control for single-frequency relaying with loop
  interference,''
\newblock {\em {IEEE} Trans. Wireless Commun.}, vol. 8, no. 6, pp. 2801--2806,
  June 2009.

\bibitem{riihonen201109TWC}
T.~Riihonen, S.~Werner, and R.~Wichman,
\newblock ``Hybrid full-duplex/half-duplex relaying with transmit power
  adaptation,''
\newblock {\em {IEEE} Trans. Wireless Commun.}, vol. 10, no. 9, pp. 3074--3085,
  Sept. 2011.

\bibitem{laneman_tse_wornell_2004}
J.N. Laneman, D.N.C. Tse, and G.W. Wornell,
\newblock ``Cooperative diversity in wireless networks: Efficient protocols and
  outage behavior,''
\newblock {\em {IEEE} Trans. Inf. Theory}, vol. 50, no. 12, pp. 3062--3080,
  Dec. 2004.

\bibitem{longversion_preprint}
M.~Khafagy, A.~Ismail, M.-S. Alouini, and S.~A\"issa,
\newblock ``On the outage performance of full-duplex selective
  decode-and-forward relaying,''
\newblock {\em {IEEE} Commun. Lett.}, vol. 17, no. 6, pp. 1180--1183, June
  2013.

\bibitem{paulraj2012WCNC}
T.~M. Kim and A.~Paulraj,
\newblock ``Outage probability of amplify-and-forward cooperation with full
  duplex relay,''
\newblock in {\em Proc. IEEE WCNC'12}, Paris, France, Apr. 2012.

\bibitem{ross-probmod}
S.~M. Ross,
\newblock {\em Introduction to Probability Models},
\newblock Academic Press, New York, 10th edition, 2010.

\end{thebibliography}
\end{document}